\documentclass[12pt,letterpaper]{article}
\usepackage{amsmath}
\usepackage{overcite}
\usepackage{graphicx}
\makeatletter
\@addtoreset{equation}{section}
\makeatother
\numberwithin{equation}{section}
\makeatletter \renewcommand\@biblabel[1]{$^{#1}$} \makeatother
\begin{document}

\title{\textbf{FIELD-THEORETIC APPROACH TO IONIC SYSTEMS: CRITICALITY AND TRICRITICALITY}}
\author{\textbf{A. Ciach}\thanks{email:\, aciach@ichf.edu.pl}
\textbf{ and G. Stell}\thanks{email:\, gstell@mail.chem.sunysb.edu}\\
\textsl{${}^{\ast}\!$Institute of Physical Chemistry, Polish Academy of Sciences},\\
\textsl{01-224 Warszawa, Kasprzaka 44/52, Poland}\\
\textsl{${}^{\dagger}\!$Department of Chemistry, State University of New York at Stony Brook},\\
\textsl{Stony Brook, NY 11794-3400, USA}}
\date{}

\maketitle

\begin{abstract}
\setlength{\baselineskip}{15pt}
A Landau-Ginzburg functional of two order  parameters (charge-density
$\phi$ and mass-density deviation  $\eta$) is developed in order to yield a 
field theory relevant to ionic lattice gases as well as a family of
off-lattice models of ionic fluids that go beyond the restricted primitive
model (RPM). In a mean-field (MF) approximation  an instability of a uniform
phase with respect to charge fluctuations with a  wave-number $k\ne 0$ is
found. This second-order transition to a charge-ordered phase terminates at  a
tricritical point (tcp). Beyond MF, a singularity of a mass correlation
function for $k\to 0$ occurs at ion concentration lower than that of the MF tcp.
An effective functional depending  only on $\eta$ is constructed. For low ion
concentration the usual Landau  form of the simple-fluid (Ising) functional
is obtained; hence in this theory the critical point is in the Ising
universality class.
\end{abstract}

\section{Introduction}\label{sec:Sec1}

Over the past few years there has been growing interest in the critical behavior
and phase separation of ionic fluids and of the Hamiltonian model that has
been most widely used in trying to understand the experimental results, the
restricted primitive model (RPM), which we shall discuss below.  The interest
stems in part from the fact that on the experimental side there has emerged no
clear picture of universality in the critical behavior of these fluids of the
sort that describes a wide variety of non-ionic fluids.  For example, in a
series of impressive experiments done on a family of ionic fluids, Narayanan
and Pitzer\cite{NarayananPitzer95} have obtained results consistent with  Ising-like
behavior.  But within the family they studied, they encountered a wide range
of reduced temperatures $t =(T-T_c)/T_c$ that mark the crossover from behavior
well-described by mean-field theory to behavior that is Ising-like as one
approaches the critical temperature $T_c$ from above along the critical
isochore.  (Some of the crossovers come at much smaller $t$ than encountered
in the case of simple non-ionic fluids.)  That means that not all members of
the family can be well-described by the RPM; whatever the crossover
temperature of the RPM, it is given by a single reduced temperature $t$
according to the law of corresponding states.
\par
Theoretical studies have revealed no simple ionic-fluid universality either. 
For example the most comprehensive
simulation\cite{{OrkoulasPanagiotopoulos96},{CaillolLevesqueWeis95},{ValleauTorrie98}} and 
theoretical\cite{{Stell92},{Fisher94},{Stell95},{LeeFisher96},{CarvalhoEvans95},{YehZhouStell96},{Stell96},{Stell99}}
studies both suggest that the critical point of the RPM\cite{{Stell99},{DickmanStell99},{HoyeStell97}} is a simple
liquid-gas critical point, whereas studies of lattice-gas versions of the
RPM\cite{{Stell99},{DickmanStell99},{HoyeStell97}} reveal that on simple cubic and body-centered cubic lattices
one should expect tricriticality instead of simple criticality.  This is in striking contrast to the situation one
finds in non-ionic fluid models, where one has come to expect off-lattice and lattice models to share the same
critical behavior. The shape of the phase diagram found in the continuous RPM from
recent Monte Carlo simulations\cite{{OrkoulasPanagiotopoulos96},{CaillolLevesqueWeis95},{ValleauTorrie98}} is
shown in Fig.1. The critical behavior revealed by these simulations is consistent with a critical point related to
the density fluctuations that represents an instability with respect to
fluctuations of wave number
$k=0$. In the case of the RPM with the positions of the ions restricted to the
lattice sites, i.e., a lattice RPM, quite different  phase behavior has been
found, both in simulation and in mean-field results\cite{{Stell99},{DickmanStell99}} as well as in a
general theoretical study.\cite{HoyeStell97}  From these references, what emerges  is  an
instability of the uniform phase induced by  charge fluctuations with wave
number $k \ne 0$. The line  of the second-order transition to the
charge-ordered phase with the period
$2\pi/k$ just below the transition ($\lambda$-line) appears to terminate at a
tricritical point, which lies in a different range of densities and 
temperatures than the critical point found in the continuum-space RPM studies,
as shown in Fig.2.
\par
In order to better understand this question of universality in ionic-systems
we have developed a field-theoretic approach for studying the boundary of
stability of a uniform phase in such systems. As prototypic microscopic 
models we consider both the lattice and the continuous versions of the RPM.
In our treatment the differences between the lattice and the continuous systems
are reflected  in different forms of the couplings in the Landau-Ginzburg (LG)
functional,  which has the same  general structure in both cases. With our
method we are able to reproduce the essential properties of ionic systems both
for the continuous-space and the lattice RPM within one formalism. We
can also  include the effects of the short-range  interactions between  various
species, i.e. extend our model beyond the RPM.   
\par
As in the earlier lattice RPM studies we have cited, an instability of the
uniform phase with respect to charge fluctuations with $k\ne 0$ is
already found at the mean-field level when we apply our approach to the
lattice RPM. We find  second-order  and  first-order transitions to a
charge-ordered phase at temperatures above and below that of a tricritical 
point respectively. The 
density at the tricritical point is found to be $1/3$ on the simplest
mean-field level for both the lattice and the continuous  versions of our LG
functional. The structure of the charge-ordered phase depends on whether  the
positions of the ions are or are not restricted to the lattice sites.

In addition to the instability induced by the charge-fluctuations with 
$k\ne 0$, an instability related to the mass-fluctuations with $k=0$ occurs 
in a different region of thermodynamic space. This other instability, which
has already been investigated on the mean-field level in the context of
several standard ionic-fluid approaches applied directly to the
 continuum-space
RPM\cite{{Stell92},{Fisher94},{Stell95},{LeeFisher96},{CarvalhoEvans95},{YehZhouStell96},{Stell96},{Stell99}} can
be detected when the  charge-fluctuations are integrated out in our formalism. We  find an equation for the
critical point consistent with  the  fluid-theory result found previously by one of us.\cite{Stell95} We also
derive the effective functional depending only on density deviations from the average value, i.e. with the
charge fluctuations integrated out.  Near the critical point this functional
has the same form as the ``$\phi^4$'' functional  representing the Ising
universality-class. Hence in our treatment of the RPM model the critical point
induced by the mass fluctuations  belongs to the Ising  universality class.

Since we find the tricritical point observed previously only in the 
lattice RPM as well as the liquid-gas type critical point found previously
only in the continuous  RPM  our approach enables us to give a unified
description of the continuum-model and lattice-model behavior of ionic
systems.   Moreover, it sharply reveals two distict
transitions that can be driven by the Coulombic forces in such models.  One is
a transition from a charge-disordered to a charge-ordered phase.  The other is
a liquid-gas type transition that does not involve charge ordering at all. In 
the continuum-space version one expects at least two solid phases
associated with different charge ordering\cite{VegaBresmaAbascal96} as
well as the fluid phase, and it may be that the locus of charge-ordering
transitions we find in our field theoretic versions of the RPM will occur in
the continuum-space RPM as a solid-solid transition,
or when one passes from a fluid to solid state.  A reliable description of
transitions involving solid phases of the continuum-space RPM that can be used
to give a global-description of both its fluid and solid phases requires going
beyond the treatment we have given here.

In lattice-gas versions of the RPM, the details of the charge ordering that
can occur will surely depend sensitively upon the lattice geometry and upon
the extent of the core of the interparticle potential excluding multiple
occupancy of neighboring sites as well as a central site as they do in
non-ionic lattice-gas models.\cite{HallStell93}  It seems likely that in some extended-core
models, one will find both the liquid-gas-like criticality and phase separation
characteristic of the continuum RPM as well as charge-ordering at higher
densities involving phases that are also density ordered and hence
solid-like.  The lattice-gas version of the RPM and its extended-core
generalization may well prove valuable in modelling the transitions one finds
in the continuum-space RPM solid.

Beyond the RPM we find evidence that the position of the tcp is extremely
sensitive to the strength of the additional, short-range interactions, as for
example the interactions between the ions and the particles of the solvent. 
For reasonable values of the parameter describing the effect of these
interactions in a simple extension of the RPM we find  within  our
formalism that the tcp  lies in the immediate vicinity of the cp found in the
recent simulations, either slightly above or slightly below the cp in
 temperature, with its
location depending very sensitively on the values of the  parameter describing
the effect of the short-range interactions.  This raises the possibility that
in some real ionic fluids in which there are substantial short-range
interaction involving the solvent, one might find a tcp realizable in the
liquid state or incipient as a tcp involving metastable liquid phases.

\section{The functional in the case of RPM (no ion-solvent interactions)}\label{sec:Sec2}

Our choice is designed to be relevant to a lattice-gas version of the RPM as
well as variants thereof, such as the standard off-lattice RPM and an
extension that includes the effect of ion-solvent interactions.
\par
In order to study criticality in ionic systems within a field-theoretic
framework, we first construct an appropriate Landau-Ginzburg (LG) 
functional.  We start the construction by calculating  the grand 
 thermodynamical potential $\Omega^{MF}$ in
the MF approximation. For the mixture  $\Omega^{MF}$ has the form
\begin{equation}\label{eqn:Eqno1}
\beta \Omega^{MF} = \beta F^h [\rho_{\alpha}({\bf r})] + 
\beta \Big( U^{MF} [\rho_{\alpha}({\bf r})] -
\int d{\bf r} \mu_{\alpha} \rho_{\alpha}({\bf r})\Big).
\end{equation}
 In the lattice case the integration in (\ref{eqn:Eqno1}) 
should be replaced by a summation over the lattice sites, with the lattice 
constant  $a$ equal to the diameter of the ions.
 $ \rho_{\alpha}$ and $\mu_{\alpha}$ are the density and the chemical
 potential of the component $\alpha$ respectively and 
 the summation convention for the repeated Greek indices is used. $U^{MF}$ is
 the energy in the MF approximation. $F^h$ is the Helmholtz free energy of 
the hard-core reference system, in the lattice (and ideal) case given by
\begin{equation}\label{eqn:Eqno2}
\beta F^h = \sum_{\bf x}\rho_{\alpha}({\bf x}) \log \rho_{\alpha}({\bf x}).
\end{equation}
In continuous models
\begin{equation}\label{eqn:Eqno3a}
F^h=\int d{\bf r} f^h[\{\rho_{\alpha}({\bf r})\}]
\end{equation}
with 
\begin{equation}\label{eqn:Eqno3b}
\beta {\partial^2f^h\over \partial\rho_{\alpha}\partial\rho_{\beta}}=
{\delta^{Kr}_{\alpha,\beta}\over \rho_{\alpha}}-c^h(\rho).
\end{equation}
Here we will limit ourselves to $F^h$ given by (\ref{eqn:Eqno2}). More accurate forms of
 $F^h$ in the continuous models will affect the explicit expressions for the
 coupling constants of the final LG functional, but not its general structure.

For the RPM model $\alpha =+,-,0$ correspond to a cation, an anion or a 
solvent particle (a `hole') respectively, and $\Omega^{MF}$ can
be expressed in terms of the local charge $\phi({\bf r})=\rho_+ -\rho_-$ and 
ion-mass $\rho({\bf r})=\rho_++\rho_-=1-\rho_0$ densities. The energy contributions to
$\Omega^{MF}$ read:

{\bf in a continuum}
\begin{equation}\label{eqn:Eqno4a}
U^{MF}-\mu_{\alpha} \int d{\bf r}\rho_{\alpha}({\bf r})=
{1\over 2}\int d{\bf r}\int_{|{\bf r}-{\bf r'}|>a}  d{\bf r'} 
V_c({\bf r}-{\bf r'})\phi({\bf r})\phi({\bf r'})
-\mu\int d{\bf r}\rho({\bf r})
\end{equation}

{\bf on a lattice}
\begin{equation}\label{eqn:Eqno4b}
U^{MF}-\mu_{\alpha} \sum_{\bf x}\rho_{\alpha}({\bf x})=
{1\over 2}\sum_{\bf x}\sum_{{\bf x'}\ne{\bf x}} 
V_c({\bf x}-{\bf x'})\phi({\bf x})\phi({\bf x'})
-\mu\sum_{\bf x}\rho({\bf x}),
\end{equation}
where $\mu=\mu_+-\mu_0=\mu_--\mu_0$. $V_c(r)=q^2/(Dr)$,
 with $q$ being the charge and $D$
 the dielectric constant, is the Coulomb potential.
 Note the restrictions on the integration in (\ref{eqn:Eqno4a}) and 
 the summation in (\ref{eqn:Eqno4b}), which are necessary to avoid the self-energy 
contributions. These restrictions have a strong effect on the form of 
$U^{MF}$ in the Fourier representation. It is convenient to introduce 
\begin{equation}\label{eqn:Eqno5}
V(r)=V_c(r) \theta (r-a).
\end{equation}
Then, with $V_c$  replaced by $V$, the integration in (\ref{eqn:Eqno4a}) and the
 summation in (\ref{eqn:Eqno4b}) are over the whole space,
 with no restrictions, and
\begin{equation}\label{eqn:Eqno6}
U^{MF} =\int {d {\bf k}\over (2\pi)^d} \tilde V(k) \tilde \phi({\bf k})
\tilde \phi(-{\bf k}).
\end{equation}
The tilde refers to the Fourier
transform of the corresponding function.
In the continuum version, an upper cutoff $\sim 2\pi/a$ is assumed as in
standard  theories, and 
in the lattice case the integral is strictly restricted to 
${\bf k}=(k_1,k_2,k_3)$ such that $-\pi\le k_i\le \pi$ 
($k_i $ measured in units $a^{-1}$). The explicit forms of
 $\tilde V(k)$ are

{\bf in the continuum}
\begin{equation}\label{eqn:Eqno7}
\tilde V({\bf k})= 4\pi {\cos k\over k^2}.
\end{equation}
The restriction on integration in (\ref{eqn:Eqno4a}), $|{\bf r}-{\bf r'}|> a$,
 results in the factor $\cos k$ 
multiplying the Fourier transform of the Coulomb potential.

{\bf on the lattice}
\begin{equation}\label{eqn:Eqno8a}
\tilde V({\bf k}) =2\pi\Bigg[ {1\over 3(1-f({\bf k}))} -V_0\Bigg]
\end{equation}
\begin{equation}\label{eqn:Eqno8b}
V_0=\int_{-\pi}^{\pi}{d k_1\over 2\pi}...
\int_{-\pi}^{\pi}{d k_3\over 2\pi} {1\over 3(1-f({\bf k}))}
\end{equation}
with
\begin{equation}\label{eqn:Eqno8c}
f({\bf k})={1\over 3}\sum_{i=1}^3 \cos k_i.
\end{equation}
The first term in (\ref{eqn:Eqno8a}) is the Fourier transform of the Coulomb potential $V_c$
 and the second term results from the fact that the summation in (\ref{eqn:Eqno4b}) is
 restricted to ${\bf x}\ne {\bf x'}$.

The $\Omega^{MF}$ assumes a minimum for the charge 
density $\phi =0$ and the  mass density $\rho =\rho^*_0$, and is next expanded
in a power series in $\phi$ and $\eta = \rho - \rho^*_0$.  Assuming that 
$\phi$ and $\eta$ are both of small magnitude and vary  slowly on the length
scale of the lattice constant or particle core diameter $a$, we truncate  the
expansion and in the lattice case use a continuous approximation for 
 $\Omega^{MF}$.

The above procedure, described in more detail elsewhere,\cite{CiachStell_ump} leads to a
functional of the two-order parameter (OP) fields, $\phi$ and $\eta$, of the following
form:
\begin{equation}\label{eqn:Eqno9a}
\Omega = \Omega_g + \Omega_{int}
\end{equation}
with the Gaussian part given by
\begin{equation}\label{eqn:Eqno9b}
\beta \Omega_g = {1\over 2!} \int d{\bf r} \left[  
a_2 \phi^2({\bf r}) + A_2\eta ^2({\bf r})\right]
 +{1\over 2!}\int d{\bf r}\int d{\bf r}' \phi({\bf r}) 
\beta V(|{\bf r-r'}|)\phi({\bf r}')
\end{equation}
and with 
\begin{equation}\label{eqn:Eqno9c}
\beta \Omega_{int} = \int d{\bf r}\left[
{1\over 3!} b_3 \phi^2({\bf r})\eta({\bf r}) + {1\over 4!} a_4 \phi^4({\bf r}) 
+ h.o.t.\right]
\end{equation}
In the above, $r$ is a dimensionless distance,  such that  $a\equiv 1$; $h.o.t.$
stands for 'higher order terms'.  In Ref.~[17] we obtain for $F^h$ given by (\ref{eqn:Eqno2}):
\begin{equation}\label{eqn:Eqno10a}
a_2 = \rho_0^{*-1}
\end{equation}
\begin{equation}\label{eqn:Eqno10b}
A_2=[\rho_0^*(1-\rho_0^*)]^{-1}
\end{equation}
\begin{equation}\label{eqn:Eqno10c}
b_3= -3\rho_0^{*-2}
\end{equation}
\begin{equation}\label{eqn:Eqno10d}
a_4=2\rho_0^{*-3}
\end{equation}
where  we use the standard notation in which $\rho^*_0$ is the dimensionless 
density $\rho a^3$ and 
\begin{equation}\label{eqn:Eqno10e}
\beta V(r)=\beta^* r^{-1} \theta(r-1)
\end{equation}
with 
$\beta^* = T^{*-1}=\beta q^2/ Da$. Higher-order couplings can also be
 expressed in terms of $\rho^*_0$.  

The functional defined in Eqs.~(\ref{eqn:Eqno9a})-(\ref{eqn:Eqno10e}) is of lowest order in $\phi$ and $\eta$, 
allowing for studying boundary of stability of the uniform phase in ionic
fluids within bifurcation analysis. Without the $h.o.t.$, however, $\Omega$ 
becomes unstable for $\phi ,\eta \to \infty$.
From Eqs~(\ref{eqn:Eqno9a})-(\ref{eqn:Eqno9c}) one can easily see that the instability occurs if 
$\Delta=(b_3^2/3-a_4A_2)/12\ge 0$ [or explicitly
 $\rho_0^*<1/3$ from Eqs.~(\ref{eqn:Eqno10a})-(\ref{eqn:Eqno10d})] and corresponds to 
$\phi,\eta \to \infty$ in such a way that $\eta= b\phi^2$ with 
$(-b_3/6-\sqrt \Delta )/A_2\le b\le (-b_3/6+\sqrt \Delta )/A_2$. For one of 
the OP fixed, $\Omega$ remains finite for the other OP going to infinity. 
When the $h.o.t.$ up to $O(\eta^4)$ are included, $\Omega$ becomes stable.
Therefore  at the later stage of our study (\ref{sec:Sec5}) we will
 also include the $h.o.t.$ up to $O(\eta^4)$, which stabilize the functional. 

In the  Fourier space $\Omega_g$ has the representation
\begin{equation}\label{eqn:Eqno11a}
\beta \Omega_g = {1\over 2} \int d{\bf k} \left(\tilde C_{\phi \phi}^0(k)
|\tilde\phi({\bf k})|^2 +A_2 |\tilde \eta ({\bf k})|^2\right)
\end{equation}
where

{\bf in the continuum}
\begin{equation}\label{eqn:Eqno11b}
\tilde C_{\phi \phi}^0(k)=\rho_0^{*-1}\left( 1 + 4\pi \beta^* \rho^*_0{\cos k
\over k^2}\right)
\end{equation}

{\bf on the lattice}
\begin{equation}\label{eqn:Eqno11c}
\tilde C_{\phi \phi}^{0 latt}(k)=\rho_0^{*-1}\left[ 1+2\pi\rho_0^*\beta^* 
 \left({1\over 3(1-f({\bf k}))} - V_0\right)\right].
\end{equation}
  As already discussed, in the lattice case the integral in (\ref{eqn:Eqno11a}) is 
restricted to $-\pi <k_i<\pi$.

In simple fluids the short-wavelength fluctuations are irrelevant in the 
critical region, and the long-wavelength fluctuations occur with the same 
probability in the continuous and in the lattice systems. This is basically
 the reason of the universality of critical phenomena.

 For ${\bf k} \to 0$ i.e. for
 distances much  larger than the lattice constant, both (\ref{eqn:Eqno11b}) and 
(\ref{eqn:Eqno11c}) are of the same form: $ Ak^{-2}
 + B$. For the long-wavelength charge fluctuations  both the lattice and the 
continuous systems behave in a similar way, as in the case of the simple 
fluids.
  Note, however that the long -
wavelength fluctuations are strongly suppressed. The Boltzmann
factor $\sim exp(-\beta\Omega[\phi,\eta])$, measuring the probability of
fluctuations $(\phi, \eta)$, approaches zero for charge fluctuations $\tilde
\phi(k\to 0)$ [see Eqs.~(\ref{eqn:Eqno11a})-(\ref{eqn:Eqno11c})].
 For a fluctuation $\tilde \phi ({\bf k})$ the charge
neutrality is violated within regions of linear size $2\pi / k$ and the 
long-wavelength fluctuations are very rare. The relevant charge fluctuations
 are  of finite wavelengths. A probability of a fluctuation 
$\tilde \phi({\bf k})$
 with $k$ finite is different in the continuous and in the lattice case 
(compare (\ref{eqn:Eqno11b}) and (\ref{eqn:Eqno11c})). 
In the RPM model the critical phenomena correspond to long-wavelength mass 
fluctuations, which are, however induced by the charge-fluctuations. Since the
 relevant charge-fluctuations are of short-wavelengths, and the probabilities 
of the  short-wavelength fluctuations in continuum and on the lattice are
 different, one may expect
 differences between the  lattice and the continuous systems.  

\section{MF results}\label{sec:Sec3}

Within the MF there is no ordinary critical point of a liquid-gas type, 
because the OPs are decoupled in the Gaussian part $\Omega_g$, and $A_2$ never
vanishes. Rather, an instability of the uniform phase with respect to {\bf
charge} fluctuations $\tilde \phi (k)$ with $k\ne 0$ occurs when  $\tilde
C_{\phi \phi}^0(k)=0$. The boundary of stability of the uniform phase 
corresponds to the highest temperature for which   $\tilde C_{\phi \phi}^0(k)$
vanishes. The bifurcation lines are

{\bf in the continuum}
\begin{equation}\label{eqn:Eqno12a}
T^*_b(\rho_0^*) ={2\pi \sin k_b\over k_b} \rho_0^*.
\end{equation}
where the wave number corresponding to bifurcation is given by
\begin{equation}\label{eqn:Eqno12b}
\tan k_b = -{2\over k_b}
\end{equation}
 from which one obtains $k_b\approx 2.46$ and the slope of the bifurcation 
line  $\approx 1.61 $.

{\bf on the lattice}
\begin{equation}\label{eqn:Eqno13a}
T^*_b=2\pi \left(V_0-{1\over6}\right)\rho_0^*
\end{equation}
and the wave vector at
 the bifurcation is
\begin{equation}\label{eqn:Eqno13b}
{\bf k}_b=(\pm \pi,\pm \pi,\pm \pi).
\end{equation}

The line of a continuous transition to a charge-ordered phase becomes first
order at a {\bf  tricritical point} (tcp). To find the position of the tcp we 
first minimize $\Omega$ with respect to $\eta$. From the condition
 $\delta\Omega/\delta\eta=0$,  we can  express $\eta$ in terms of
 $\phi$,
\begin{equation}\label{eqn:Eqno14}
 \eta({\bf r})= \eta_0({\bf r})
= - {b_3\over 3!A_2}  \phi^2({\bf r})
\end{equation}
 and in the vicinity of the bifurcation we have
\begin{eqnarray}
\beta \Omega\!\! &=&\!\! {1\over 2} \int {d{\bf k}\over (2\pi)^d} 
\tilde C_{\phi \phi}^0(k)
|\tilde\phi({\bf k})|^2\nonumber \\
& & \\ \label{eqn:Eqno15}
&+& {{\cal A}_4\over 4!} 
\int {d{\bf k}_1 \over (2\pi)^d}\int {d{\bf k}_2\over (2\pi)^d}
 \int {d{\bf k}_3 \over (2\pi)^d}\int{ d{\bf k}_4\over (2\pi)^d}
(2\pi)^d\delta(\sum_{i=1}^4 {\bf k}_i)\prod_{i=1}^4\tilde\phi({\bf k}_i) +
 O(\epsilon^6)\nonumber
\end{eqnarray}
where 
\begin{equation}\label{eqn:Eqno16}
{\cal A}_4 =a_4-{b_3^2\over 3 A_2}
=3\rho_0^{*-3}\left(\rho_0^*-{1\over 3}\right),
\end{equation}
 $\epsilon$ is the bifurcation parameter,  $\epsilon \sim
\sqrt{|T^*-T_b^*|/T^*_b}$, such that 
 $\tilde C_{\phi \phi}^0(k)=O(\epsilon^2)$.
At local minima of $\Omega$ ( $\delta\Omega/\delta\phi=0$ and 
$\delta\Omega/\delta\eta=0$)
$\phi=O(\epsilon)$ and $\eta=O(\epsilon^2)$. 
From the above we find that the  tcp occurs at
 $\rho_{0t}^*=1/3$ for the model
given by Eqs.~(\ref{eqn:Eqno9a})-(\ref{eqn:Eqno10e}). 

 Just below the bifurcation various metastable structures can be described 
as linear combinations of planar waves
\begin{equation}\label{eqn:Eqno17}
\phi({\bf r})= \Phi \cos ( {\bf k}_b\cdot {\bf r})
\end{equation}
where $\hat {\bf k}_b$ is a direction of oscillations and the amplitude
 is $\Phi=O(\epsilon)$. 

{\bf on the lattice} 

${\bf r}=(r_1,r_2,r_3)$ are restricted to integer $r_i$.  For the integer $r_i$
(\ref{eqn:Eqno17}) and (\ref{eqn:Eqno13b})  give $\phi({\bf r})= \pm \Phi$. Thus two sublattices, one
positively- and the other one  negatively charged occur.  A charge-ordered phase
with such a structure has been also found directly in the lattice RPM
in MF results and in Monte Carlo simulations.\cite{{Stell99},{DickmanStell99}}

{\bf in the continuum}

To determine the structure stable just below the continuous transition, one
 has to find for what linear combinations of planar waves (\ref{eqn:Eqno17})
  $\Omega$ assumes the global minimum.
 For example, the structure of ionic
 crystal (like NaCl) corresponds to a superposition of four waves, with the
 four vectors  $\hat {\bf k}_b$ forming a tetrahedron.\cite{CiachStell_ump}
In Ref.~[17] we verify by explicit calculations of $\Omega$ approximated by (\ref{eqn:Eqno15})
 and (\ref{eqn:Eqno16}) that for a single wave (\ref{eqn:Eqno17}), i.e. for a  lamellar 
structure,  $\Omega$ assumes a lower value than for the other structures.
 Formation of charged layers, with the
 neighboring layers  oppositely charged, is energetically favorable 
compared to the disordered phase when the average distance between 
alike ions within one layer is larger than the distance between
 oppositely charged layers. The 
ratio between the  average distance between the ions within one layer  and 
the distance between oppositely charged layers can be estimated on the basis 
of $\rho_0^*$ and the period of modulations $\lambda =2\pi/k_b\approx
2.55 a$. Near the tcp this 
ratio is  $\ge \sqrt {(2/\lambda)^3/\rho_0^*} \approx 1.2$.  

We should
stress here that for $a\to 0$ (or with $\theta(r-1)$ in (\ref{eqn:Eqno10e}) omitted), i.e.
for point charges, no such instability takes place, because then 
$\tilde C_{\phi
\phi}^0(k)>0$, and consistently, $\lambda \to 0$ for $a\to 0$. Note also that
the tcp lies far away from the  critical point (cp), for which the
latest simulation estimate is\cite{{OrkoulasPanagiotopoulos96},{CaillolLevesqueWeis95},{ValleauTorrie98}} 
$(\rho_c^*,T^*_c)\approx (0.08,0.05)$.

The charge-charge correlation function in the MF approximation, $\tilde
G^0_{\phi \phi}= \tilde C^{0 -1}_{\phi \phi}$ , never  diverges for $k\to 0$,
rather,  $\tilde G^0_{\phi \phi}\to 0$ for $k\to 0$ (see (\ref{eqn:Eqno11b}) and (\ref{eqn:Eqno11c})),
 i.e. $\int d
{\bf r} G_{\phi\phi}^0(r)=0$.  By inspection of Eqs.~(\ref{eqn:Eqno11b}) and~(\ref{eqn:Eqno11c})
 for $k\to 0$ we can easily
identify the inverse Debye screening length $\kappa a = \sqrt{ 4\pi \rho^*_0
\beta^*}$, describing the decay of the charge-charge correlations.

\section{Critical mass fluctuations}\label{sec:Sec4}

Although there is no `ordinary' criticality within the MF, the charge 
fluctuations may lead to the critical behavior of the mass-mass correlation
function for $k\to 0$, because the two OP-s are coupled beyond the Gaussian 
part of $\Omega$.   Fluid
theories\cite{{Stell92},{Fisher94},{Stell95},{LeeFisher96},{CarvalhoEvans95},{YehZhouStell96},{Stell96},{Stell99}}
and  simulations\cite{{OrkoulasPanagiotopoulos96},{CaillolLevesqueWeis95},{ValleauTorrie98}} predict low $\rho^*$
at critical; thus we concentrate 
on 
$\rho_0^* < \rho^*_{0t}=1/3$.

In order to determine whether the criticality of the liquid-gas type occurs 
in this model, we calculate the correlation function for the mass fluctuations
$G_{\eta \eta}$, defined as
\begin{equation}\label{eqn:Eqno18}
G_{\eta \eta}({\bf r}_1,{\bf r}_2)=\langle \eta({\bf r}_1 ) \eta({\bf r}_2)
\rangle - \langle  \eta({\bf r}_1) \rangle  \langle  \eta({\bf r}_2 )\rangle ,
\end{equation}
in standard perturbation expansion about the Gaussian solution.
  Next we examine
the behavior of the Fourier transform of $G_{\eta \eta}$  for $k\to 0$,
searching for singular behavior.

The most probable mass fluctuation $\eta_0({\bf r})$, 
accompanying a given charge fluctuation 
$\phi({\bf r})$, is given by (\ref{eqn:Eqno14}) (if the $h.o.t.$ in Eqs.~(\ref{eqn:Eqno9a})-(\ref{eqn:Eqno9c}) are
included, $\eta_0({\bf r})$ contains additional
contributions of a form $\sim \phi^{2n}$; see Ref.~[17]).
Arbitrary mass fluctuation can be written as
 $\eta=\eta_0+\Delta\eta = -{b_3\over 3!A_2}\phi^2+\Delta\eta$.
 The functional $\Omega[\phi,\eta]$ given by Eqs.~(\ref{eqn:Eqno9a})-(\ref{eqn:Eqno9c}) 
can be expanded about the minimum at $\eta_0$. Then one obtains a functional of
 $\phi,\Delta \eta$ of the form
\begin{equation}\label{eqn:Eqno19}
\beta\Omega[\phi,\Delta\eta] = \beta\Omega_{aux} [\phi] + 
{1\over 2}A_2 \int d{\bf r}(\Delta\eta({\bf r}))^2 +O(\Delta \eta ^4).
\end{equation}
The $\Omega_{aux}[\phi]$
 is given by (\ref{eqn:Eqno15}), or in the real space representation by 
\begin{equation}\label{eqn:Eqno20}
\beta \Omega _{aux} =\int d{\bf r}\left( 
{1\over 2}\int d {\bf r'}  C_{\phi \phi}^0(|{\bf r}-{\bf r'}|)\phi({\bf r})
\phi({\bf r'}) +
{1\over 4!}{\cal A}_4 \phi^4({\bf r})+ O(\phi^6)\right)
\end{equation}
 It is convenient to introduce the functional (\ref{eqn:Eqno19}),
because $ A_2$ is just a constant in this model and 
$\Delta \eta({\bf r})$ contribute only to local parts of the mass-mass 
correlation functions. The nonlocal correlations, relevant in the critical
 region, are determined by the 
auxiliary functional $\Omega_{aux}$ (\ref{eqn:Eqno20}), depending only on the single OP 
$\phi$. 

In this description the  one- and n-point correlation - functions for mass
fluctuations can be expressed in terms of average values of the field
$\phi^2({\bf r})$ and of a  product of $\phi^2$ at n points, respectively. The
explicit relations   for the nonlocal parts of the correlation functions read
\begin{equation}\label{eqn:Eqno21a}
\langle \eta({\bf r})\rangle 
= - {b_3\over 3!A_2} \langle \phi^2({\bf r})\rangle _{aux}
\end{equation}
\begin{equation}\label{eqn:Eqno21b}
G_{\eta \eta}^{nl}({\bf r}_1,{\bf r}_2)= 
\left({b_3\over 3!A_2}\right)^2
 G_{\phi^2 \phi^2}({\bf r}_1, {\bf r}_2)
\end{equation}
where 
\begin{equation}\label{eqn:Eqno21c}
 G_{\phi^2 \phi^2} ({\bf r}_1, {\bf r}_2)=\langle 
(\phi^2({\bf r}_1) -\langle \phi^2({\bf r}_1)\rangle _{aux})
(\phi^2({\bf r}_2) -\langle \phi^2({\bf r}_2)\rangle _{aux})\rangle _{aux} 
\end{equation}
and
\begin{equation}\label{eqn:Eqno22}
G_{n\eta}^{nl}({\bf r}_1,...,{\bf r}_n) = 
\left( {b_3\over 3!A_2}\right)^n 
G_{n\phi^2 }({\bf r}_1,..., {\bf r}_n)
\end{equation}
In the above $G_{n\phi^2 }$ is defined in a way similar to the definition of  
$G_{\phi^2 \phi^2 }$ in Eq.~(\ref{eqn:Eqno21c}). $\langle ...\rangle_{aux}$ means averaging with
the Boltzmann factor $\sim \exp (-\beta \Omega_{aux})$.

In order to find the critical point we  consider $G_{\phi^2 \phi^2}$, which is
just proportional to $G_{\eta \eta}^{nl}$, in the perturbation expansion, and
verify whether there occur singular contributions for $k\to 0$.  $G_{\phi^2
\phi^2}({\bf r}_1, {\bf r}_2)$ can be expressed, in the standard way,\cite{Amit84}
in terms of Feynmann diagrams of the connected four-point  function, with the
vertex ${\cal A}_4$ and the free-propagator $G^0_{\phi \phi}$, in which two
pairs of external points are identified with each other. Next  the Fourier
transform for every diagram is  calculated.  Because $\tilde G^0_{\phi
\phi}(k)$ is regular for $k \to 0$, and even tends to zero (see Eqs.(\ref{eqn:Eqno11b})
 and (\ref{eqn:Eqno11c})), every individual diagram is
regular for $k \to 0$; hence, any finite sum of diagrams is regular as well.

From Eq.~(\ref{eqn:Eqno16}) we see that the effective coupling constant ${\cal A}_4$ is 
{\bf negative } for $\rho_0^*<1/3$, i.e. for densities lower than the tcp
density. The negative ${\cal A}_4$ is of crucial  importance for the  behavior
of $\tilde G_{\phi^2 \phi^2}$. For illustration consider a Fourier transform 
of a series of chains of loops, with the n-th order term (n vertices ${\cal
A}_4$ and n+1 loops ) given by
\begin{equation}\label{eqn:Eqno23a}
\alpha _n(k) = (-{\cal A}_4 )^n \tilde g_0^{n+1}(k)
\end{equation}
with 
\begin{equation}\label{eqn:Eqno23b} 
\tilde g_0(k)= {1\over 2}\int d {\bf r} G_{\phi \phi}^{0}(r)^2 
e^{i {\bf k r}}={1\over 2}\int d {\bf k'}\tilde G_{\phi\phi}^0 ({\bf k'})
\tilde G_{\phi\phi}^0 ({\bf k-k'}).
\end{equation}

A contribution to  $\tilde G_{\phi^2 \phi^2}$ given by the sum of such diagrams
has a form
\begin{equation}\label{eqn:Eqno24}
\sum_0^{\infty} \alpha _n(k) = \tilde g_0(k) 
\left[ 1+ {\cal A}_4\tilde g_0(k)\right]^{-1}.
\end{equation}
In the part of $(\rho_0^*,T^*)$ to which we restrict our attention here,
$\tilde g_0(k)$ is a smooth function (see (\ref{eqn:Eqno23b}) and (\ref{eqn:Eqno11b}), (\ref{eqn:Eqno11c})). 
 For ${\cal A}_4<0$, however, (\ref{eqn:Eqno24}) represents a  singular contribution to 
$G_{\eta \eta}^{nl}$
 for $k\to 0$  if 
\begin{equation}\label{eqn:Eqno25}
1+{\cal A}_4 \tilde g_0(0)=0.
\end{equation}
The above equation is the lowest order approximation for the critical
singularity. A more accurate equation has a form
\begin{equation}\label{eqn:Eqno26}
1+{\cal A}_4 \tilde g(0)=0,
\end{equation}
where  $\tilde g(k)$ is a Fourier transform of the function  representing a sum
of all connected diagrams of four-point functions with  two  pairs of
external points identified, which cannot be split into two distinct diagrams by
splitting a single hyper-vertex ${\cal A}_4$ into two two-point vertices.  In
other words, $g(r)$ has no contributions which are of a form of chains; $g_0$
is the lowest order approximation for $g$. For $\rho_0^*\to 0$ and with $g({\bf
r})$ approximated by $G_{\phi \phi}^2({\bf r})/2$, where $G_{\phi \phi}$ is
 the charge-charge correlation function, (\ref{eqn:Eqno26}) reduces to an equation
obtained previously by one of us within fluid theory.\cite{Stell95}

By summing up a related geometric series (as in Eqs.(\ref{eqn:Eqno23a}) and (\ref{eqn:Eqno24}),
 with $g_0$
replaced by $g$) we obtain  a more accurate form of the  singular part of
$G_{\eta \eta}^{nl}$, which we will denote by $\tilde G^0_{\eta \eta}$,
\begin{equation}\label{eqn:Eqno27}
\tilde G^0_{\eta \eta}(k) =  \tilde g(k) 
\left[ 1+ {\cal A}_4\tilde g(k)\right]^{-1}.
\end{equation}

In the critical region we can expand $\tilde C^0_{\eta \eta} (k) = \tilde
G^0_{\eta \eta} (k)^{-1}$ about $k=0$ and we obtain
\begin{equation}\label{eqn:Eqno28}
\tilde C^0_{\eta \eta} (k)= \alpha_0 + \alpha_2 k^2
\end{equation}
with $\alpha_0$ and $\alpha_2$ expressed in terms of $\tilde g(0)$ and $\tilde
g(0)''$ in a  standard way.

Consider the higher-order mass - correlation functions $G_{n \eta}^{nl}$ and 
related vertex functions $\Gamma_{n\eta}$ for $n\ge 3$ and with the two-point
function approximated by (\ref{eqn:Eqno27})
\begin{equation}\label{eqn:Eqno29}
G_{n\eta }^{nl}({\bf r}_1,..., {\bf r}_n)= 
- \int d{\bf r}'... \int d{\bf r}^{n}
\Gamma_{n\eta}( {\bf r}',... , {\bf r}^n) \times 
G_{\eta \eta}^0 ({\bf r}_1, {\bf r}') ...
G_{\eta \eta}^0 ({\bf r}_n, {\bf r}^n)
\end{equation}
Unless the terms $\sim \phi^6,\sim \phi^8$ are explicitly included in (\ref{eqn:Eqno9c}),
there are no contributions to $\Gamma_{n\eta}$ at the zero-loop level.  At the
one-loop approximation, for $\Gamma_{3\eta}$, for example, we have
\begin{equation}\label{eqn:Eqno30}
\Gamma_{3\eta}( {\bf r}_1,{\bf r}_2 , {\bf r}_3) =
G_{\phi \phi}^0({\bf r}_1, {\bf r}_2)
G_{\phi \phi}^0({\bf r}_2, {\bf r}_3)G_{\phi \phi}^0({\bf r}_3, {\bf r}_1)
\cdot  {\rm num. fac. + O(2-loop)}
\end{equation}
and similarly for $n>3$. After integration (\ref{eqn:Eqno30}) yields
\begin{equation}\label{eqn:Eqno31}
\Gamma_{n} = \int d{\bf r}_1 ...d{\bf r}_n \Gamma_{n\eta}({\bf r}_1, ..., 
{\bf r}_n) \sim \tilde G_{\phi \phi}^0 (0)^n =0
\end{equation}
Higher-order contributions to $\Gamma_n$ also contain a factor $\tilde G_{\phi
\phi} (0) =0$. Hence, for $n\ge 3$ $\Gamma_n$ give  negligible contribution to
the effective potential.\cite{Amit84}

In the perturbation expansion for the charge-charge correlations 
$\tilde G_{\phi \phi}(k)$,
 every Feynmann diagram  in the Fourier representation contains  a factor 
$\tilde  G_{\phi \phi}^{0n} (k)$ with $n\ge 1$, which for $k=0$ vanishes. 
In an infinite series of diagrams, similar to those which lead to divergent
 mass correlations, a segment of a corresponding chain  for 
$\tilde G_{\phi \phi}$ 
is proportional to $\tilde G_{\phi \phi}^0(k)$. The 
infinite series of Feynmann diagrams gives a contribution to 
charge-charge correlations which
 vanishes for $k\to 0$ (compare (\ref{eqn:Eqno24}) with $\tilde g_0(k)$ replaced by a term 
proportional to
 $ \tilde G_{\phi \phi}^0(k)$). Hence the 
charge-correlation length remains finite at the critical point, at which the
 mass-correlation length diverges. This is consistent with the results found in our
earlier liquid-state theoretic analysis.\cite{{Stell92},{Stell95},{Stell99}}

\section{Construction of the effective functional for $\rho_0^*<1/3$}\label{sec:Sec5}

In order to construct an appropriate effective functional depending only
on $\eta$ in the critical region, we recall that $\Omega$ given by (\ref{eqn:Eqno9a})
without the $h.o.t.$ is  unstable for $\phi, \eta \to \infty $ with 
$\eta\sim\phi^2$ if
$\rho_0^*<1/3$ .  We thus include the terms stabilizing $\Omega$ and we
consider the functional
\begin{equation}\label{eqn:Eqno32a}
{\cal F}[\phi,\eta] =\Omega[\phi, \eta] + \Omega_{\eta}[\eta]
\end{equation}
with
\begin{equation}\label{eqn:Eqno32b}
\Omega_{\eta} [\eta]= \int d{\bf r} \left( {A_3\over 3!} \eta({\bf r})^3
+{A_4\over 4!} \eta({\bf r})^4\right)
\end{equation}
where, by construction of the functional\cite{CiachStell_ump}
\begin{equation}\label{eqn:Eqno32c}
A_3={1\over (1-\rho_0^*)^2} - {1\over \rho_0^{*2}}
\end{equation}
\begin{equation}\label{eqn:Eqno32d}
A_4=2\left[{1\over (1-\rho_0^*)^3} + {1\over \rho_0^{*3}}\right]
\end{equation}
if $F^h$ is approximated by (\ref{eqn:Eqno2}).

The effective functional of $\eta$ is formally defined by
\begin{equation}\label{eqn:Eqno33}
e^{-\beta \Omega_{eff}[\eta]}=\int D\phi e^{-\beta {\cal F}[\phi,\eta]}=
e^{-\beta \Omega_{\eta}[\eta]} \int D\phi e^{-\beta \Omega[\phi, \eta]}.
\end{equation}
Thus,
\begin{equation}\label{eqn:Eqno34a}
\Omega_{eff}[\eta] = \Omega_{\eta}[\eta] +{\cal F}_0[\eta]
\end{equation}
with
\begin{equation}\label{eqn:Eqno34b}
e^{-\beta {\cal F}_0 [\eta]}=\int D\phi e^{-\beta \Omega[\phi, \eta]}
\end{equation}

It is not possible to perform the functional integration in (\ref{eqn:Eqno34b}) exactly, but
it is also not necessary for the purpose of characterizing the
mass-correlation singularity. We only need an approximate functional, which 
leads to the same behavior of the correlation functions at large distances in 
the critical region that the full
${\cal F}$ does. The above-described analysis  of the critical behavior of the
correlation functions leads to
\begin{equation}\label{eqn:Eqno35}
\beta {\cal F}_0 [\eta]= \int d{\bf k} \left[ 
{\tilde C_{\eta \eta}^0 (k) \over 2} \delta\eta({\bf k})\delta \eta(-{\bf k})
\right] + {\rm corr.}
\end{equation}
where $\tilde C_{\eta \eta}^0 (k)$ is given by (\ref{eqn:Eqno28}) and 
\begin{equation}\label{eqn:Eqno36}
\delta\eta=\eta-\langle\eta\rangle
\end{equation}
with $\langle\eta\rangle$ given by (\ref{eqn:Eqno21a}).
With the above form of ${\cal F}_0$  the two-point mass correlation function
(\ref{eqn:Eqno27}) is recovered and $\langle \rho\rangle =\rho^* +\langle\eta\rangle$
 Terms $\sim \eta^n$ 
with $n\ge 3$ are neglected in (\ref{eqn:Eqno35}). As usual, for small amplitude
fluctuations such terms are negligible compared to terms $\sim \eta ^2$,
except for the critical fluctuations for which the second-order term vanishes
at the critical point. For such  fluctuations, i.e. $\eta({\bf r})=const$,
however, the higher-order  contributions to ${\cal F}_0$ vanish (unless the
 terms $O(\phi^6),O(\phi^8), O(\phi^4\eta), O(\phi^2\eta^2) $ etc.
 are explicitly included in $\Omega[\phi,\eta]$, 
see (\ref{eqn:Eqno31})).

By shifting and rescaling the field,
\begin{equation}\label{eqn:Eqno37}
\psi ({\bf r}) = \left( \eta({\bf r}) + {A_3\over A_4}\right) \sqrt{\alpha_2}
\end{equation}
we obtain from (\ref{eqn:Eqno34a}), (\ref{eqn:Eqno32b}) and (\ref{eqn:Eqno35}) a familiar form of the Ising
universality-class functional, namely
\begin{equation}\label{eqn:Eqno38a}
\Omega _{eff}[\psi] = \int d{\bf r}\Bigl[ -h \psi ({\bf r}) +{u_2\over 2} 
\psi({\bf r}) ^2 + {1\over 2}(\nabla \psi ({\bf r}))^2 
+ {u_4\over 4!} \psi ({\bf r})^4\Bigr] + h.o.t.
\end{equation}
with explicit expressions for the couplings corresponding to the functional
 (\ref{eqn:Eqno32b}) and (\ref{eqn:Eqno9a}) given by
\begin{equation}\label{eqn:Eqno38b}
h={A_3\over A_4}\left( \alpha_0 -{ A_3^2\over 3A_4}\right)
\alpha_2^{-1/2}
\end{equation}
\begin{equation}\label{eqn:Eqno38c}
u_2=\left( \alpha_0 - {A_3^2\over 2 A_4}\right) \alpha_2^{-1}
\end{equation}
and
\begin{equation}\label{eqn:Eqno38d}
u_4=A_4\alpha_2^{-2}
\end{equation}
Higher-order contributions to (\ref{eqn:Eqno32a}) as well as more accurate approximation for 
the hard-core reference system (Eqs.~(\ref{eqn:Eqno3a}) and~(\ref{eqn:Eqno3b}) instead of (\ref{eqn:Eqno2})) modify
the couplings. In particular, to
 $A_n$ ($n=3,4$) additional terms, related to the coupling constants of the 
terms $\sim\eta^k\phi^{2n-k}$ should be added, if the latter are included in 
the functional (\ref{eqn:Eqno32a}). Such terms have significant effect on the position of the
 critical point. In the MF approximation for the functional (\ref{eqn:Eqno38a})  the cp is 
given by $ h(\rho_0^*,\beta^*)=u_2(\rho_0^*,\beta^*)=0$, which reduce to 
$A_3=\alpha_0=0$. The density of the critical point, given by $A_3=0$, changes 
significantly when $A_3$ is replaced by a sum of several terms depending
 on $\rho^*$ resulting from the 'h.o.t.'. However, the 'h.o.t.'  
 have no qualitative effect on the universality class.

The $u_2$ depends on $\beta^*$ through $\alpha_0$ and $\alpha_2$, 
which in turn are functions of $\tilde g(0)$ and  $\tilde g(0)''$ (see Eq.(\ref{eqn:Eqno28}) 
and below). The function $\tilde g(k)$ is determined by the charge 
fluctuations, and for $k\to 0$ depends on the whole spectrum of the 
fluctuations, as one can see from the lowest-order approximation for 
$\tilde g$, $\tilde g_0$, given by Eq.(\ref{eqn:Eqno23b}). Because of that dependence 
on the shape of 
$\tilde G_{\phi\phi}^0(k)$, the 
position of the cp, $u_2=0$, also depends on the form of
 $\tilde G_{\phi\phi}^0(k)$.
 In particular, for 
different systems, characterized in our theory by different forms of
 $\tilde G_{\phi\phi}^0$ (as, for example, the continuous (\ref{eqn:Eqno11b}) and the
 lattice (\ref{eqn:Eqno11c}) models) the relative locations of a possible cp signaling an
Ising-like transition and a possible locus of transitions to a charge-ordered
phase can be quite different.  As a result, one or the other of these
transitions can be preempted by the occurrence of the other.  Thus the
continuum-space RPM fluid appears to support only the cp while the cubic
lattice-gas RPM appears to support only the charge-ordering transition with an
attendant tcp.  It will require analysis beyond that given here to locate the
cp in our field-theoretic approach.

\section{Beyond the RPM (short-range interactions included)}\label{sec:Sec6}

Here we limit ourselves to the continuous case. The lattice case is described in Ref.~[17].
On our level of description the short-range interactions, including those between the ions
and the solvent, can be represented by terms $a_2 \xi_{\phi}^2(\nabla \phi)^2/2$ and  $A_2
\xi_{\eta}^2(\nabla\eta)^2/2$ as in standard Landau theories for short-range
interactions. In  general, $a_2$ and $A_2$  differ, respectively, from  those of
Eqs.(\ref{eqn:Eqno10a}) and~(\ref{eqn:Eqno10b}),
 which refer  to the  absence of the short-range interactions.
The $\xi_{\phi} $ and $\xi_{\eta}$ are the (MF) spin and mass correlation lengths,
respectively,  in a system with the Coulomb interactions turned off. They
should be thus comparable to the typical distance between the ions and the
particles of  the solvent. For our length unit both should be thus of order of
unity.  Here we restrict ourselves to the case of $a_2$ and $A_2$ given by 
(\ref{eqn:Eqno10a}) and (\ref{eqn:Eqno10b}) and to a single length $\xi=\xi_{\phi}=\xi_{\eta}$, to 
illustrate on the simplest level the crucial role of the interactions
 beyond the RPM on the position
 of the $\lambda$-line and the tricritical point. 

The bifurcation analysis of
the  extended functional (\ref{eqn:Eqno9a}), which in addition contains  the terms  
$a_2 \xi^2(\nabla \phi)^2/2$ and  $A_2 \xi^2(\nabla 
\eta)^2/2$, gives the 
 generalization of Eqs.~(\ref{eqn:Eqno12a}) and~(\ref{eqn:Eqno12b}) for the
bifurcation line
\begin{equation}\label{eqn:Eqno39a}
\tan k_b=-2 {1+2\xi^2k_b^2\over k_b(1+\xi^2k_b^2)}
\end{equation}
and
\begin{equation}\label{eqn:Eqno39b}
 T^*_b(\rho_0^*)=2{\pi\sin k_b\over k_b(1+2\xi^2k_b^2)}\rho_0^*.
\end{equation}
The tcp is very sensitive to the changes of $\xi$ and is given by the
analysis of Ref.~[17] as
\begin{equation}\label{eqn:Eqno40}
\rho^*_{tcp}={1\over 3+8\xi^2k_b^2}.
\end{equation}
For $\xi=1$, for example, $\rho_{tcp}^*(\xi=1)= 0.026$ and 
$T^*_{tcp}(\xi=1)=0.007$, which is well below the cp (compare the
position of the tcp in the absence of the ion-solvent interactions,
$\rho_{tcp}^*(\xi=0)=1/3$ and $T^*_{tcp}(\xi=0)=0.54$).

In the case of ions large compared to the particles of the solvent, as for
example in the system $N_{2226}B_{2226}$ studied by several groups\cite{SinghPitzer90} a 
reasonable estimate for the typical distance between the ions and the particles
of the solvent should be around 0.5.  From (\ref{eqn:Eqno40}) and Eqs.~(\ref{eqn:Eqno39a}) and~(\ref{eqn:Eqno39b})
we find  the tcp for two slightly different values of $\xi$,  $\xi=0.5$  and
$\xi=0.55$ at $(\rho^*,T^*)=(0.08,0.055)$ and $(0.07,0.043)$
respectively. Note that in the first case the tcp lies slightly above the RPM
cp=$(0.08,0.05)$, actually found in the
simulations,\cite{{OrkoulasPanagiotopoulos96},{CaillolLevesqueWeis95},{ValleauTorrie98}}  whereas in the second
case the tcp lies below the observed cp.

It is important to note the approximations and assumptions we are making in
using Eqs.~(\ref{eqn:Eqno39a}), (\ref{eqn:Eqno39b}), and (\ref{eqn:Eqno40}) as we do.  There are precise values of
$\xi_{\phi}$ and
$\xi_{\eta}$ associated with each thermodynamic state in our system and, in
particular, with the states along the bifurcation line associated with (\ref{eqn:Eqno39a}), (\ref{eqn:Eqno39b}), 
and the tcp associated with (\ref{eqn:Eqno40}).  We do not know these precise values,
however, so we make simple estimates of their magnitudes and then see how
sensitive the tcp $T^*$ and $\rho^*$ that come out of (\ref{eqn:Eqno39a}), (\ref{eqn:Eqno39b}), and
(\ref{eqn:Eqno40}) are to slight changes in $\xi$.  We find great sensitivity, which suggests
that the tcp location is very sensitive to the interaction parameters upon
which the $\xi_{\phi}$ and $\xi_{\eta}$ depend.

Whether a similar sensitivity of the cp location upon these parameter values
exists is an important question that we cannot probe on the basis of our
results here but which is deserving of further work.  Such a sensitivity
might help explain the remarkable disparities that different groups\cite{SinghPitzer90}
have found in the cp location in $N_{2226}B_{2226}$ samples.  An even more
dramatic possibility is raised by the observation that comes out of our work
that depending on the details of the short-range interactions in an ionic system
one might have both a cp and a tcp [as in Fig. 3] or only one of these
occuring in stable equilibrium, with the other describing metastable
criticality that lies within the coexistence region defined by the stable
singular point [as in Fig. 4].  We report further details in Ref.~[17].

\vspace{16pt}

\noindent{\bf{\Large Acknowledgments}}

\vspace{16pt}

A. Ciach would like to thank Prof. M.M. Telo da Gama and Dr. A.G. Moreira for very intersting discussions.
The work of A. Ciach was partially supported by a KBN grant 3 T09A 07316 and partially
supported by the National Science Foundation.  G. Stell's work was supported by
the Division of Chemical Sciences, Office of Basic Energy Sciences, Office of
Science, U.S. Department of Energy.

\newpage
\noindent{\large{\bf Figure captions}}

\vspace{16pt}

\noindent Fig.~1.- Liquid-gas coexistence curve  of the continuous RPM model
according to recent simulation
studies.\cite{{OrkoulasPanagiotopoulos96},{CaillolLevesqueWeis95},{ValleauTorrie98}} Here $\rho_0^*$ is the 
dimensionless concentration of ions and $T^*$ the reduced temperature,
 defined after Eq.~(\ref{eqn:Eqno10e}).

\vspace{12pt}
\noindent Fig.~2.- (a) $\lambda$-line and coexistence curve from cell 
histograms taken from Monte Carlo simulation results for the lattice RPM 
desribed in Ref.~[12]. (b)  $\lambda$-line and coexistence curve from the 
simple mean-field theory described in the Appendix of Ref.~[12] for the same 
lattice RPM. (Note the difference in the temperature scales.)

\vspace{12pt}

\noindent Fig.~3.-Generic phase-diagram associated with our field-theoretic
 description of the RPM, shown schematically.  Here $\rho $ and $T$ denote the
ion-concentration and  temperature respectively  in arbitrary units.
We show the case in which both a critical point and a tricritical point are
 realized, along with a $\lambda$-line (dashed) representing a locus of 
transitions to a charge-ordered state lying below the $\lambda$-line in 
temperature.

\vspace{12pt}

\noindent Fig.~4.- Possible  phase-diagrams associated with our field-theoretic
 description of the RPM with additional short-range interactions 
$a_2 \xi^2(\nabla \phi)^2/2$ and  $A_2 \xi^2(\nabla \eta)^2/2$ included,\break 
shown schematically.  Here  $\rho $ and $T$ denote the ion-concentration and 
temperature respectively  in arbitrary units. We show the two special cases
 described in the text:
(a) the critical point is realized only as a metastable point. (b) the 
tricritical point is realized only as a metastable point.

\vfil\eject
\newpage

\includegraphics[scale=0.7,angle=90]{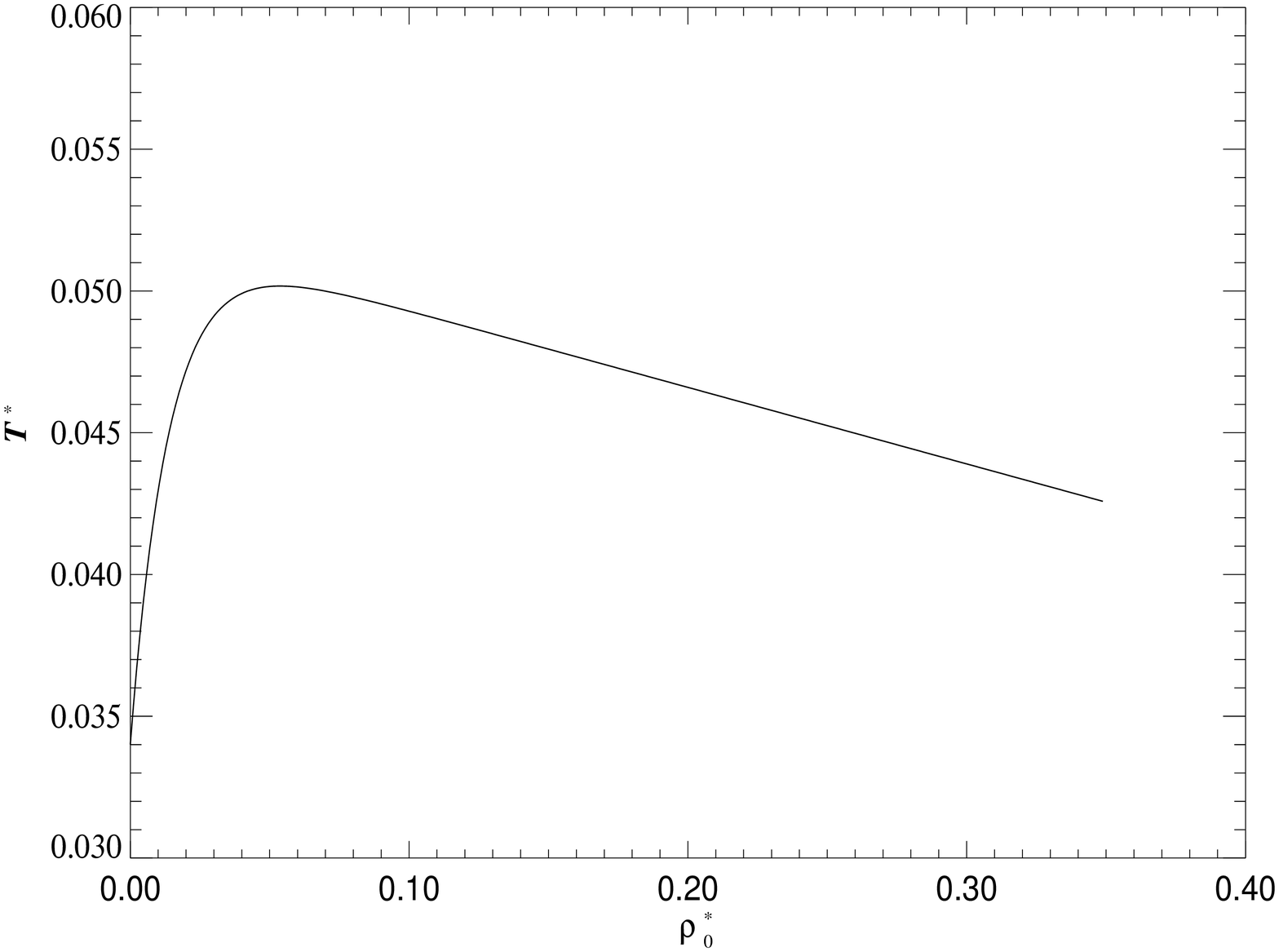}
\vspace{1cm}
\center{Ciach-Stell, Fig. 1}
\newpage

\includegraphics[scale=0.7,angle=90]{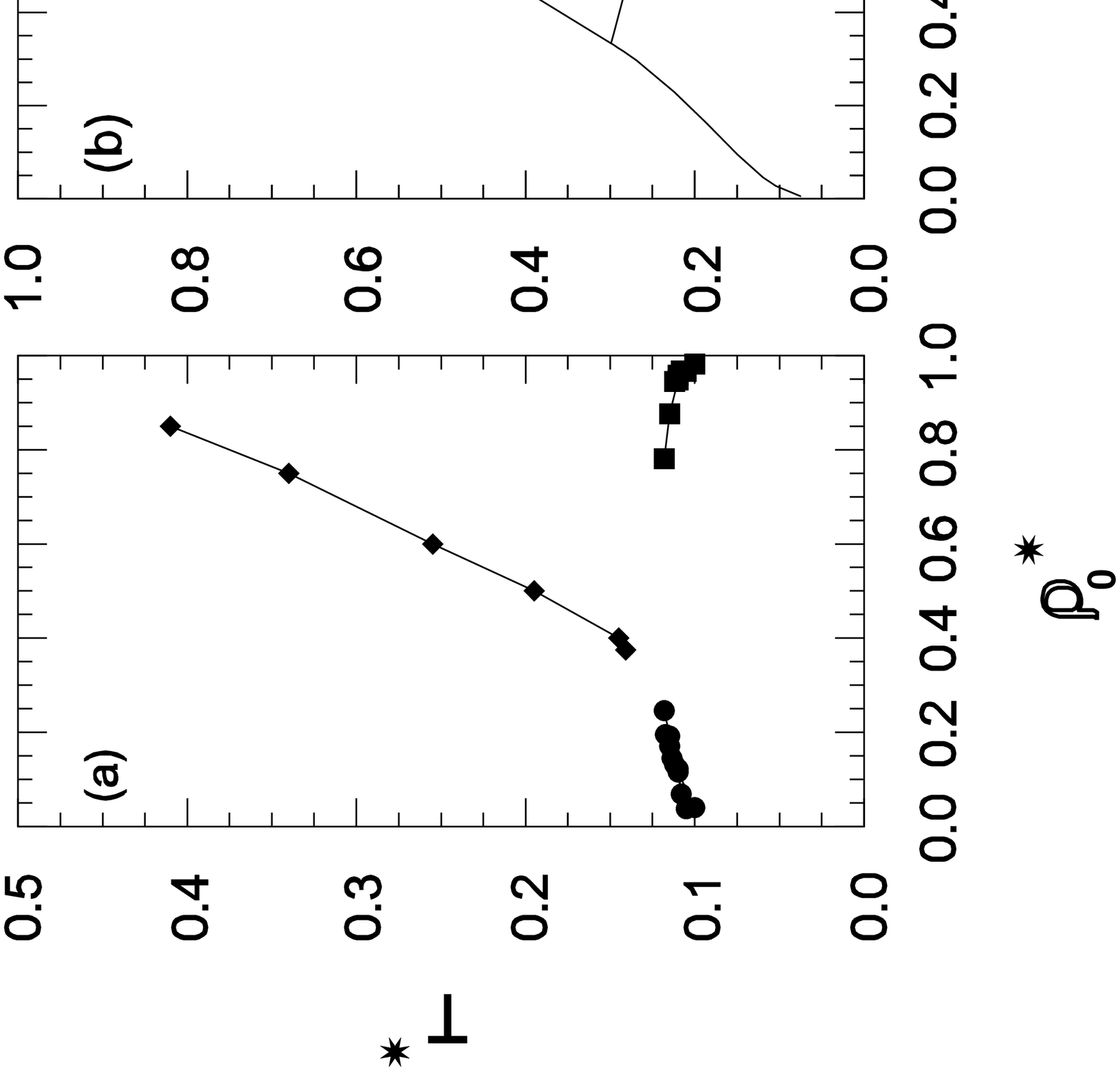}
\vspace{1cm}
\center{Ciach-Stell, Fig. 2}

\newpage

\includegraphics[scale=0.7,angle=90]{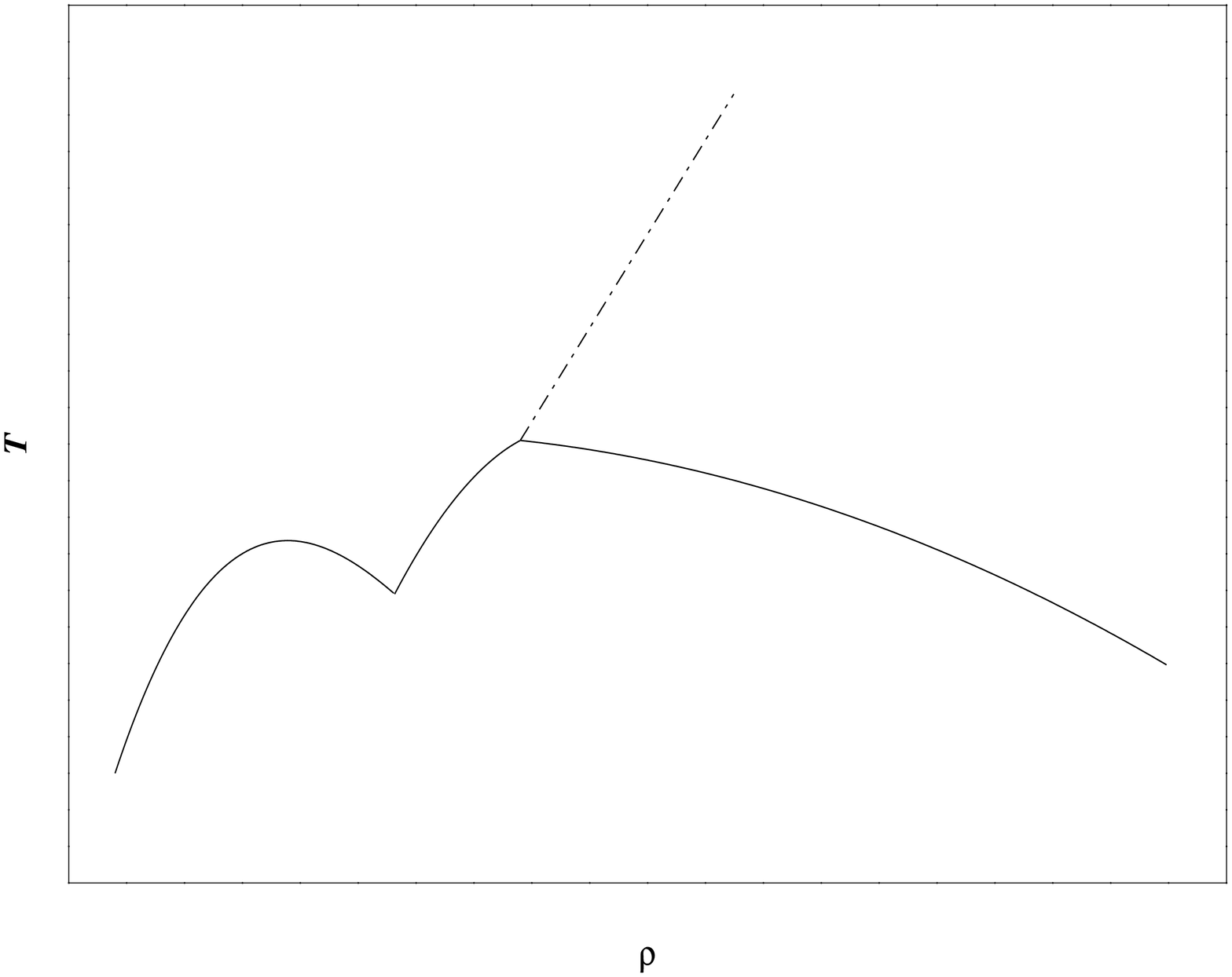}
\vspace{1cm}
\center{Ciach-Stell, Fig. 3}

\newpage

\includegraphics[scale=0.7,angle=90]{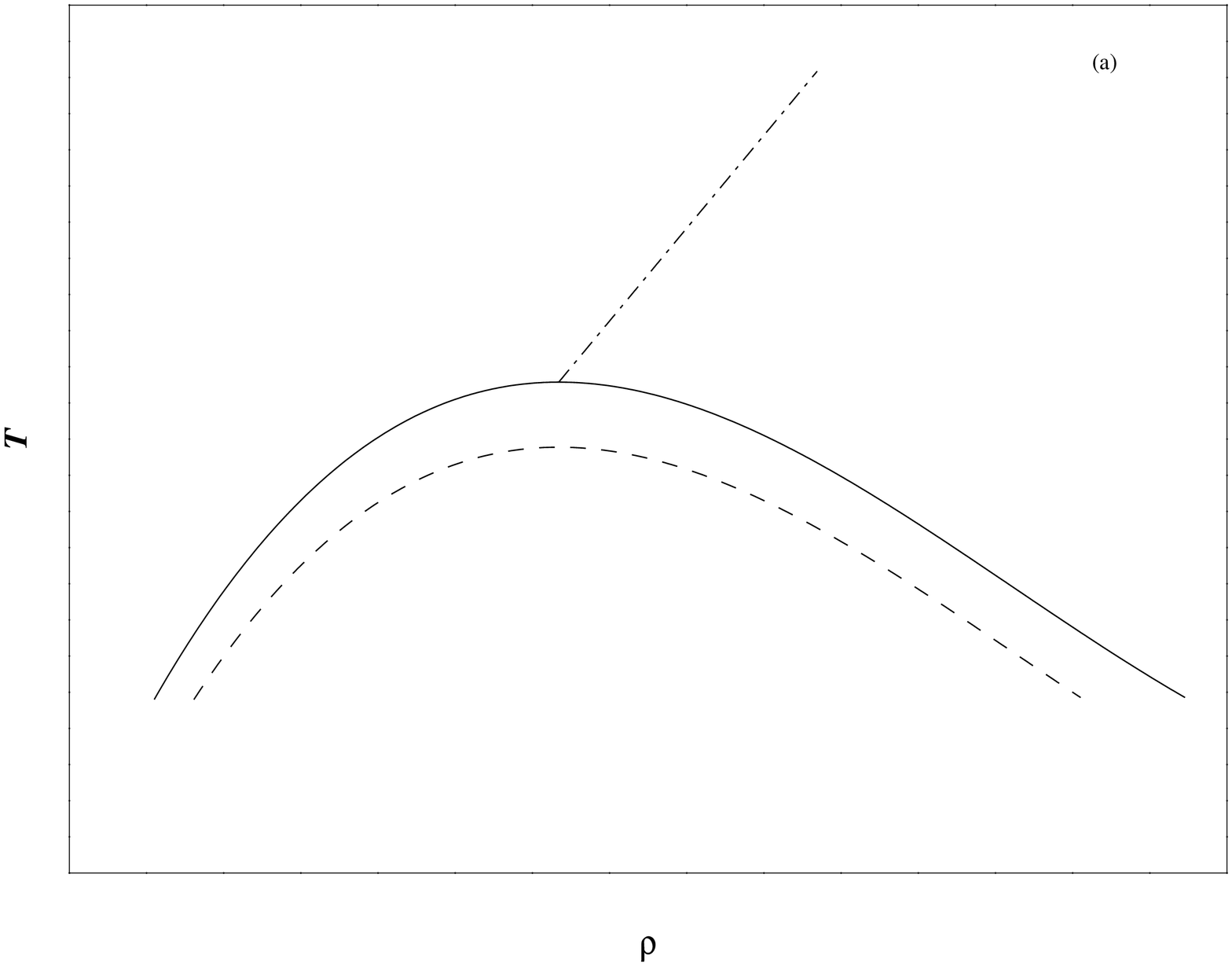}
\vspace{1cm}
\center{Ciach-Stell, Fig. 4(a)}

\newpage

\includegraphics[scale=0.7,angle=90]{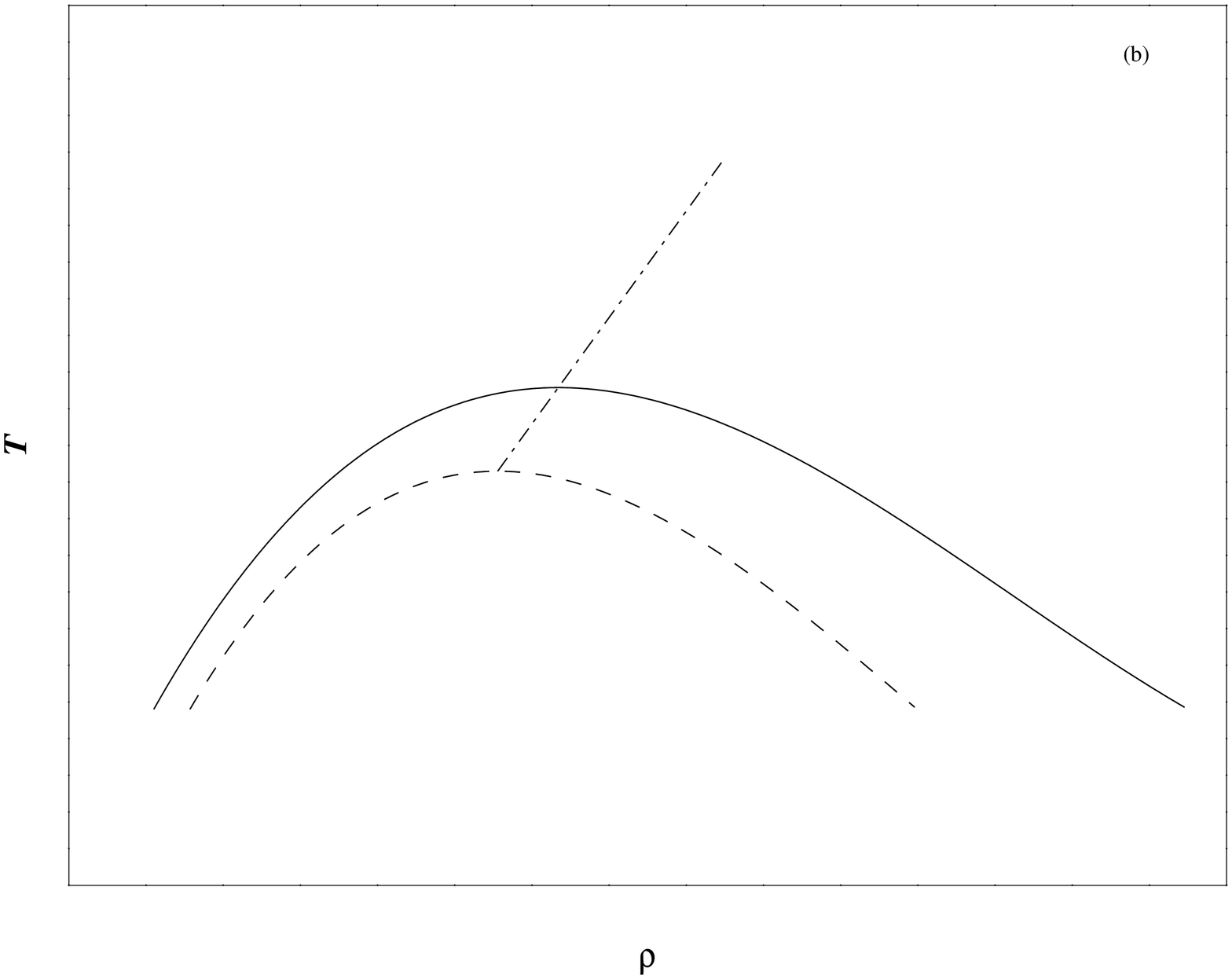}
\vspace{1cm}
\center{Ciach-Stell, Fig. 4(b)}

\end{document}